\documentclass[showpacs,aps,graphicx,twocolumn]{revtex4}
\usepackage{amssymb}
\usepackage{amsmath}%preprint,
\usepackage{graphicx}
\usepackage{multirow}

\begin{document}

\title{Circuit QED: Cross-Kerr-effect induced by a superconducting qutrit without classical pulses}
\author{Tong Liu$^1$, Chui-ping Yang$^2$, Yang Zhang$^1$, Chang-shui Yu$^1$\footnote{Corresponding author: ycs@dlut.edu.cn}, and Wei-ning Zhang$^1$}

\address{$^1$School of Physics and Optoelectronic Technology, Dalian University of Technology, Dalian 116024, P.R. China}
\address{$^2$Department of Physics, Hangzhou Normal University, Hangzhou, Zhejiang 310036, China}
\date{\today}

\begin{abstract}
The realization of cross-Kerr nonlinearity is an important task for
many applications in quantum information processing.
In this work, we propose a method for realizing cross-Kerr nonlinearity interaction between two superconducting coplanar waveguide resonators coupled by a three-level superconducting flux qutrit (coupler). By employing the qutrit-resonator dispersive interaction, we derive an effective Hamiltonian involving two-photon number operators and a coupler operator. This Hamiltonian can be used to describe a cross-Kerr nonlinearity interaction between two resonators when the coupler is in the ground state. Because the coupler is unexcited during the entire process, the effect of coupler decoherence can be greatly minimized. More importantly, compared with the previous proposals, our proposal does not require classical pulses. Furthermore, due to use of only a three-level qutrit, the
experimental setup is much simplified when compared with previous proposals requiring a four-level artificial atomic systems.
Based on our Hamiltonian, we implement a two-resonator qubits controlled-phase gate and generate a two-resonator entangled coherent state. Numerical simulation shows that
the high-fidelity implementation of the
phase gate and creation of the entangled coherent state are feasible with current circuit
QED technology.
\end{abstract}

\pacs{03.67.Lx, 42.50.Pq, 85.25.Cp, 42.50.Dv}\maketitle

\section{Introduction}
\label{sec1}
Circuit quantum electrodynamics (QED), consisting of
microwave resonators and superconducting (SC) qubits, is an analogue of cavity QED and a well-established platform  for the investigation of light-matter
interaction at the quantum level \cite{04Blais,s2,s3,s4}. SC qubits are promising candidates to achieve
scalable quantum computing and quantum information processing (QIP), especially with continuing improvements
to coherence times \cite{s6,s7,s8,s9,s10,s11,s12,Pop,s13}. Furthermore, improving the quality factor of SC resonators is a key
development for QIP. For example, a SC coplanar waveguide
resonators with a loaded quality factor $Q\sim 10^{6}$ \cite{s40,s41} or with
internal quality factors above one million ($Q>10^{7}$)
have been previously reported \cite{s42}. Recently, a SC microwave resonator
with a loaded quality factor $Q\sim 3.5\times10^{7}$ has been
demonstrated in experiments \cite{s43}. Combined with the large electric dipole moment of SC qubit,
the strong-coupling or ultrastrong-coupling regime of a SC qubit with a resonator
has been reported in experiments \cite{04Wallraff,04Chiorescu,10Forniaz,10Niemczyk}.

The microwave resonators have been
considered as good memory elements in QIP \cite{s43,M. Mariantoni}. Based on circuit QED, there is much interest in large-scale QIP which usually involves two or more microwave resonators/cavities.
Recently, a number of proposals have been presented for synthesizing entangled states
(e.g., Bell states, GHZ states, NOON states, and entangled coherent states) of two or more than two resonators \cite{08Mariantoni, 10Strauch, njpMerkel,11Hu, 12Strauch, 12yang, 13yang, 14Su,15Xiong, 16Sharma, 16Zhao,Z. Li}, and realizing two-qubit or multiqubit quantum gates with microwave photons distributed in two resonators \cite{11Strauch,M. Hua}.
In addition, a great deal of quantum
effects and operations involved multiple microwave resonators have been experimentally demonstrated in circuit QED. For instance,
Ref.~\cite{M. Mariantoni} experimentally implemented the quantum von Neumann architecture with
SC circuits, Ref.~\cite{H. Wang} generated photon NOON states of two SC
resonators, Ref.~\cite{L. Steffen} realized full deterministic
quantum teleportation with feed-forward,
and Ref.~\cite{C. Wang} created a two-mode
cat state of microwave fields in two SC cavities, respectively.
Those progresses in circuit QED provide a
promising perspective of microwave photons as resource for quantum communication
and computation.

Cross-Kerr nonlinearity, known as cross phase modulation, is one of the most
promising tool for quantum computation and QIP. In quantum optics, when two
photons are simultaneously input into a nonlinear medium, the output photons
undergo nonlinear optical effects named the cross-Kerr nonlinearity effect. During the last several decades, many QIP tasks have been proposed in optical systems based on cross-Kerr nonlinearity, including construction of quantum phase gates \cite{Q. A. Turchette,K. Nemoto}, generation of macroscopic quantum superposition states \cite{C. C. Gerry,H. Jeong}, completion of quantum teleportation \cite{D. Vitali}, realization of the
quantum-nondemolition (QND) measurements \cite{N. Imoto,P. Grangier}, and implementation of entanglement purification \cite{Y. B. Sheng}.

On the other hand, many theories and experiments involved cross-Kerr effect have been discussed in circuit QED \cite{S. Rebi,O. Suchoi,S. Kumar,F. R. Ong,G. Kirchmair,B. Fan,I. Hoi,Y. Hu,E. T. Holland,H. Zhang}. Experimentally, based on the cross-Kerr nonlinearity, the observation of quantum state collapse and revival \cite{G. Kirchmair}, the investigation of the feasibility of
microwave photon counting \cite{B. Fan}, and the realization of the giant cross-Kerr effect for propagating microwaves \cite{I. Hoi} have been reported.
In recent years, the cross-Kerr nonlinearity between two microwave resonators has been extensively researched
in circuit QED \cite{Y. Hu,H. Zhang,E. T. Holland}. For example, Refs.~\cite{Y. Hu,H. Zhang} proposed a scheme for implementing cross-Kerr nonlinearity between two SC resonators via an $N$-type SC artificial atomic systems, and Ref.~\cite{E. T. Holland} experimentally demonstrated a state
dependent shift between two microwave cavities via a cross-Kerr effect. The $N$-type four-level nature atoms was studied in Refs.~\cite{H. Schmidt,H. Kang,Y. F. Chen}, and the Refs.~\cite{Y. Hu,H. Zhang} extended it to the SC circuits  which consists of two
SC transmon qubits coupled by a SQUID (quantum interference device).
The Hamiltonian for a cross-Kerr interaction between
two resonators is given by (in units of $\hbar =1$)
\begin{eqnarray}
H_{\chi}=-\chi_{ab} \hat{n}_a \hat{n}_b,
\end{eqnarray}
where $\chi_{ab}$ is the coupling strength and $\hat{n}_a$ ($\hat{n}_b$) is the photon number
operator of resonator $a$ ($b$).

In this paper, we propose a method to realize a cross-Kerr nonlinearity interaction between two microwave resonators by coupling a three-level SC artificial atom~[Fig.~1(a)]. This proposal has the following features and advantages: (\romannumeral1) Different from the previous works \cite{Y. Hu,H. Zhang},
in our scheme only one operational step is needed, only one single three-level qutrit is used, and
no need to use classical pulses. Thus the operation and experimental setup are greatly simplified. (\romannumeral2)
Because resonator photons are virtually excited and the qutrit is unexcited during the entire process,
the effect of resonator decay, the unwanted inter-resonator crosstalk, and the qutrit decoherence are greatly minimized.
(\romannumeral3) This proposal can be applied to accomplish the same
task with various SC qutrits (e.g., SC charge qutrits, transmon qutrits, Xmon
qutrits, phase qutrits) coupled to
two 1D resonators or two 3D cavities.
(\romannumeral4) Numerical simulation shows that our cross-Kerr interaction Hamiltonian can be used to high-fidelity realize a
two-resonator qubits controlled-phase gate and generate a two-resonator entangled coherent state.

This paper is organized as follows. In Sec. \ref{sec2}, we show how to
realize a cross-Kerr interaction effect between two SC resonators in circuit QED. In Sec. \ref{sec3},
we show how to use our effective Hamiltonian
to construct a controlled-phase gate on two resonators, and then discuss
how to create a macroscopic entangled
coherent state of two resonators.
In Sec. \ref{sec4}, we discuss the possible experimental implementation of our
proposal and numerically calculate the operational fidelity for realizing a controlled-phase gate and generating
a entangled coherent state. A concluding summary is given
in Sec. \ref{sec5}.

\section{Cross-Kerr nonlinearity effect in circuit QED} \label{sec2}

\begin{figure}[tbp]
\begin{center}
\includegraphics[bb=66 590 421 754, width=9.0 cm, clip]{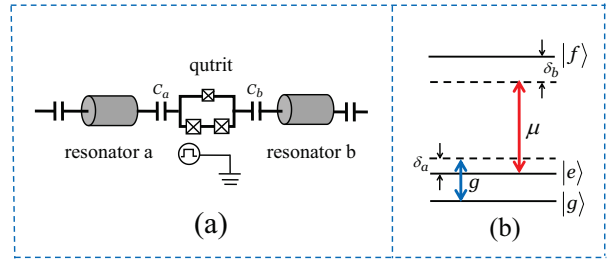} \vspace*{%
-0.08in}
\end{center}
\caption{(color online)
(a) Setup of two SC resonators coupled to a qutrit (coupler)
via capacitances $C_a$ and $C_b$. The
flux qutrits can be other types of solid-state qutrit, such as a quantum dot, a SC phase qutrit, a charge qutrit or a transmon qutrit.
(b) Resonator $a$ is far-off resonant with $|g\rangle\leftrightarrow|e\rangle$ transition of coupler with
coupling strength $g$ and detuning $\delta_1$, while
resonator $b$ is far-off resonant with $|e\rangle\leftrightarrow|f\rangle$ transition of coupler with
coupling strength $\mu$ and detuning $\delta_2$. Here, $\delta_a~(\delta_a=\omega_{eg}-\omega_{a}<0)$ is a negative detuning but $\delta_b$ ($\delta_b=\omega_{fe}-\omega_{b}>0)$ is a positive detuning. }
\label{fig:1}
\end{figure}

Consider a system consists of two SC coplanar waveguide resonators connected by a ladder-type SC flux qutrit (coupler) [Fig.~1(a)]. As shown in Fig.~1 (a,b), resonator $a$ ($b$) is off-resonantly coupled to the $|g\rangle\leftrightarrow|e\rangle $ ($|e\rangle\leftrightarrow|f\rangle $) transition of qutrit with a coupling constant $g$ ($\mu$).
In the interaction picture, the
Hamiltonian of the whole system can be written as (in units of $\hbar =1$)
\begin{eqnarray}
H_{I}=g(e^{i\delta_a t}a\sigma
_{eg}^{+}+h.c.)+ \mu(e^{i\delta_b t}b\sigma_{fe}^{+}+h.c.),
\end{eqnarray}
where $\sigma
_{eg}^{+}=|e\rangle\langle g|$ and $\sigma
_{fe}^{+}=|f\rangle\langle e|$, $\delta_a$ ($\delta_a=\omega_{eg}-\omega_{a}<0$) is a negative detuning but $\delta_b$ ($\delta_b=\omega_{fe}-\omega_{b}>0)$ is a positive detuning. Here, $\omega_{fe}$ ($\omega_{eg}$) is the $|e\rangle\leftrightarrow|f\rangle $ ($|g\rangle\leftrightarrow|e\rangle $) transition frequency of qutrit and $\omega_{a}$ ($\omega_{b}$) is the frequency of resonator $a$ ($b$).

Consider the large-detuning conditions $\delta_a\gg g$ and $\delta_b\gg\mu$, the Hamiltonian (2) becomes \cite{s35}
\begin{eqnarray}
H_{e} &=&-\frac{g^2}{\delta_a}(a^\dagger a|g\rangle\langle g|- aa^\dagger|e\rangle\langle e|)\nonumber \\
&-&\frac{\mu^2}{\delta_b}(b^\dagger b|e\rangle\langle e|-b b^\dagger|f\rangle\langle f|)\nonumber \\
&+&\lambda(e^{-i\Delta t}a^\dagger b^\dagger\sigma_{fg}^{-}+ h.c.),
\end{eqnarray}%
where $\lambda=\frac{g\mu}{2}(\frac{1}{|\delta_a|}+\frac{1}{\delta_b})$, $\Delta=\delta_b-|\delta_a|$, and $\sigma
_{fg}^{-}=|g\rangle\langle f|$.
Under the large-detuning conditions $\Delta\geq\{g^2/|\delta_a|, \mu^2/\delta_b, \lambda\}$ and $\delta_b>|\delta_a|$, the effective Hamiltonian $H_e$ changes to
\begin{eqnarray}
H_{e} &=&-\frac{g^2}{\delta_a}(a^\dagger a|g\rangle\langle g|- aa^\dagger|e\rangle\langle e|)\nonumber \\
&-&\frac{\mu^2}{\delta_b}(b^\dagger b|e\rangle\langle e|-b b^\dagger |f\rangle\langle f|)\nonumber \\
&+&\chi(aa^\dagger bb^\dagger|f\rangle\langle f|-a^\dagger a b^\dagger b|g\rangle\langle g|),
\end{eqnarray}%
where $\chi$ ($\chi=\lambda^2/\Delta$) is the cross-Kerr interaction coefficient. When the levels $|e\rangle$ and $|f\rangle$ are not occupied, the effective Hamiltonian (4) reduces to
\begin{eqnarray}
H_{e}&=&H_{0}+H_{i}
\end{eqnarray}
with
\begin{eqnarray}
H_{0}&=&-\frac{g^2}{\delta_a}a^\dagger a|g\rangle\langle g|=-\frac{g^2}{\delta_a}\hat{n}_a|g\rangle\langle g|, \nonumber \\
H_{i}&=&-\chi a^\dagger a b^\dagger b|g\rangle\langle g|=-\chi \hat{n}_a \hat{n}_b|g\rangle\langle g|,
\end{eqnarray}
where $\hat{n}_a$ and $\hat{n}_b$ are
the photon number operators for resonators $a$ and $b$. When the qutrit is in the state $|g\rangle$, the Hamiltonian $H_{0}$ describes the photon-number-dependent shift of the
resonator $a$, the Hamiltonian $H_{i}$ describes the cross-Kerr photon-photon interaction between
the resonators $a$ and $b$.

From Eq.~(6), one can see that $[H_{0},H_{i}]=0$, thus, we obtain the effective cross-Kerr interaction Hamiltonian
\begin{eqnarray}
\widetilde{H}_{e}=e^{iH_0t}H_ie^{-iH_0t}=-\chi \hat{n}_a \hat{n}_b|g\rangle\langle g|.
\end{eqnarray}
It should be noted that the Hamiltonian (7) is different from the well-known cross-Kerr Hamiltonian
$H_{\chi}=-\chi_{ab} \hat{n}_a \hat{n}_b$ describing the cross-Kerr interaction of two resonators with
coefficient $\chi_{ab}$. It is because that the Hamiltonian (7) contains a coupler operator
$|g\rangle\langle g|$ which is not involved in $H_{\chi}$. In the next section, we first show how to use the effective Hamiltonian (7) to
construct a controlled-phase gate on two resonators, and then discuss
how to use the effective Hamiltonian (7) to create a macroscopic entangled coherent state of two resonators.

\section{Controlled-phase gate implementation and entangled-state preparation}
\label{sec3}

Assume that the resonator $a$ ($b$) is in
an arbitrary pure state $|\varphi\rangle_a$ ($|\varphi\rangle_b$) and the coupler is in the state $|g\rangle$.
In this section, we first show how to use the Hamiltonian (5) or (7) to construct a two-qubit controlled-phase gate of two resonators $a$ and $b$. We then discuss how to generate a macroscopic entangled coherent state for two resonators with Hamiltonian (5) or (7).

\subsection {Controlled-phase gate on two resonators}
\label{sec31}

Suppose that the resonators $a$ and $b$ are in an arbitrary superposition
state, respectively. Assume that $|\varphi\rangle_a=\alpha|0\rangle_a+\beta|1\rangle_a$ and $|\varphi\rangle_b=\gamma|0\rangle_b+\delta|1\rangle_b$.
Here, $\alpha,$ $\beta$, $\gamma$, and $\delta$ are the normalized complex numbers; $|0\rangle_a$ or $|1\rangle_a$ ($|0\rangle_b$ or $|1\rangle_b$) represents the vacuum state or the single photon state of resonator $a$ ($b$). The time-evolution operator for the Hamiltonian (5) is defined as $U=\exp {[i(g^2 /\delta_a)\hat{n}_a t}]\otimes\exp({i\chi \hat{n}_a \hat{n}_b t})$. Therefore, under the Hamiltonian (5), the resonator-system
state $|\varphi\rangle_a$$|\varphi\rangle_b$ evolves into
\begin{eqnarray}
&&U|\varphi\rangle_a|\varphi\rangle_b\nonumber \\
&=&e^{i(g^2 /\delta_a)\hat{n}_a t}e^{i\chi \hat{n}_a \hat{n}_b t}|\varphi\rangle_a|\varphi\rangle_b\nonumber \\
&=&e^{i(g^2 /\delta_a)\hat{n}_a t}e^{i\chi \hat{n}_a \hat{n}_b t}(\alpha\gamma|0\rangle_a|0\rangle_b
+\alpha\delta|0\rangle_a|1\rangle_b\nonumber \\
&+&\beta\gamma|1\rangle_a|0\rangle_b
+\beta\delta|1\rangle_a|1\rangle_b)\nonumber \\
&=&\alpha\gamma|0\rangle_a|0\rangle_b+\alpha\delta|0\rangle_a|1\rangle_b\nonumber \\
&+&e^{i(g^2 /\delta_a)t}\beta\gamma|1\rangle_a|0\rangle_b
+e^{i\theta t}\beta\delta|1\rangle_a|1\rangle_b,
\end{eqnarray}
where $\theta=\chi+(g^2 /\delta_a)$. If we choose the interaction time $t=\pi/|\theta|$, one can obtain a two-resonator qubits controlled-phase
gate
\begin{eqnarray}\begin{split}
|&0\rangle_a|0\rangle_b\rightarrow |0\rangle_a|0\rangle_b,~~~|0\rangle_a|1\rangle_b\rightarrow |0\rangle_a|1\rangle_b,\\
|&1\rangle_a|0\rangle_b\rightarrow|1\rangle_a|0\rangle_b,~~~|1\rangle_a|1\rangle_b\rightarrow-|1\rangle_a|1\rangle_b,
\end{split}\end{eqnarray}
where we have set $(g^2 /|\delta_a|)t=2k\pi$ ($k$ is a positive integer). For $t=\pi/|\theta|$ and $(g^2 /|\delta_a|)t=2k\pi$, one has the following relationship between the parameters
\begin{eqnarray}
\lambda=g\sqrt{\frac {(\delta_b-|\delta_a|)(2k-1)}{2k|\delta_a|}}, \nonumber \\
\mu=\frac {2|\delta_a|\delta_b}{|\delta_a|+\delta_b}\sqrt{\frac {(\delta_b-|\delta_a|)(2k-1)}{2k|\delta_a|}}.
\end{eqnarray}
Notice that the effective coupling strength $\lambda$ and the coupling strength $\mu$ can be adjusted
by varying the detuning $\delta_a$ or $\delta_b$. The level
spacings of the SC ``artificial atom" can be rapidly adjusted by varying
external control parameters (e.g., magnetic flux applied to the
superconducting loop of a superconducting phase, transmon, Xmon or flux
qubit; see, e.g.,~\cite{s8,08M. Neeley,09P. J. Leek,13J. D. Strand}).

The controlled-phase gate of Eq.~(9) is one of the paradigmatic gates for quantum information and quantum computation. By using it
with single-qubit gates, a set of universal gates can be constructed. Hitherto, a great deal of theoretical proposals have been presented for realizing two-qubit controlled-phase gate in many physical systems. In circuit QED, a two SC qubits controlled-phase gate has already been experimentally demonstrated \cite{12Fedorov,s45,s46,s47}. In addition, the controlled-phase gate with two microwave-photon-resonator qubits has been previously
proposed in Refs.~\cite{11Strauch,M. Hua}. The proposals \cite{11Strauch,M. Hua}
require several operational steps and the application
of classical pulses. Compared with Refs.~\cite{11Strauch,M. Hua}, our proposal is much improved
because our phase-gate can
be achieved only using a single-step operation and no pulses are needed.

\subsection{Creation of a two-resonator macroscopic entangled coherent state}
\label{sec32}

Assume that the resonators $a$ and $b$ are in coherent states $|\varphi\rangle_a=|\alpha_a\rangle$ and $|\varphi\rangle_b=|\beta_b\rangle$, respectively.
Under the Hamiltonian (7), the joint state $|\varphi\rangle_a|\varphi\rangle_b$ of the resonators evolves into
\begin{small}
\begin{eqnarray}
|\psi\rangle&=&e^{i\chi \hat{n}_a \hat{n}_bt}|\varphi\rangle_a|\varphi\rangle_b\nonumber \\
&=&e^{-\frac{1}{2}|\alpha_a^2|}e^{-\frac{1}{2}|\beta_b^2|}\sum_{n_a,n_b=0}^{\infty}
\frac{\alpha_a^{n_a}\beta_b^{n_b}\exp{(i\chi n_a n_bt)}}{\sqrt{n_a!n_b!}}|n_a\rangle |n_b\rangle.\nonumber \\
\end{eqnarray}
\end{small}
When the evolution time is equal to $t=\pi/\chi$, one has $\exp{(i\chi n_a n_bt)}=(-1)^{n_an_b}$. We divide the sum of Eq. (11) into a part with $n_a$ even and another with
$n_a$ odd.
It is apparent that an even/odd coherent state of $|\alpha_a\rangle$ can be expressed as
\begin{eqnarray}\begin{split}
&e^{-\frac{1}{2}|\alpha_a^2|}\sum_{n_a=\textmd{even}}^{\infty}\frac{\alpha_a^{n_a}}{\sqrt{n_a!}}|n_a\rangle \\
=&e^{-\frac{1}{2}|\alpha_a|^2}(|0\rangle+\frac{\alpha_a^2}{\sqrt{2!}}|2\rangle+\cdots
+\frac{\alpha_a^{n_{a}}}{\sqrt{n_{a}!}}|n_a\rangle)  \\
=&\frac{1}{2}(|\alpha_a\rangle+|-\alpha_a\rangle),
\end{split}\end{eqnarray}
\begin{equation}\begin{split}
&e^{-\frac{1}{2}|\alpha_a^2|}\sum_{n_a=\textmd{odd}}^{\infty}\frac{\alpha_a^{n_a}}{\sqrt{n_a!}}|n_a\rangle \\
=&e^{-\frac{1}{2}|\alpha_a|^2}(\alpha_a|1\rangle+\frac{\alpha_a^3}{\sqrt{3!}}|3\rangle+\cdots+
\frac{\alpha_a^{n_a}}{\sqrt{n_a!}}|n_a\rangle) \\
=&\frac{1}{2}(|\alpha_a\rangle-|-\alpha_a\rangle).
\end{split}\end{equation}
On substituting Eqs.~(12) and~(13) into Eq.~(11), one obtains a macroscopic entangled coherent state
\begin{small}
\begin{eqnarray}\begin{split}
|\psi\rangle&
=e^{-\frac{1}{2}|\alpha_a^2|}e^{-\frac{1}{2}|\beta_b^2|}\left \{\sum_{n_a=\textmd{even}}^{\infty}\right .\frac{\alpha_a^{n_a}}{\sqrt{n_a!}}|n_a\rangle
\sum_{n_b=0}^{\infty}\frac{\beta_b^{n_b}}{\sqrt{n_b!}}|n_b\rangle\\
&+\sum_{n_a=\textmd{odd}}^{\infty}\frac{\alpha_a^{n_a}}{\sqrt{n_a!}}|n_a\rangle
\left .\sum_{n_b=0}^{\infty}\frac{(-\beta_b)^{n_b}}{\sqrt{n_b!}}|n_b\rangle  \right \}  \\
=\frac{1}{2}&(\left|\alpha_a\right\rangle+\left|-\alpha_a\right\rangle)\left|\beta_b\right\rangle+\frac{1}{2}(\left|\alpha_a\right\rangle-
\left|-\alpha_a\right\rangle)\left|-\beta_b\right\rangle \\
=\frac{1}{2}&(\left|\alpha_a\right\rangle\left|\beta_b\right\rangle+\left|-\alpha_a\right\rangle
\left|\beta_b\right\rangle+\left|\alpha_a\right\rangle\left|-\beta_b\right\rangle-\left|-\alpha_a\right\rangle\left|-\beta_b\right\rangle).
\end{split}\end{eqnarray}
\end{small}
After returning to the original interaction picture by performing a unitary transformation $U=\exp {[i(g^2 /\delta_a)\hat{n}_a t}]$,
the entangled coherent state (14) becomes
\begin{small}
\begin{eqnarray}
|\varphi\rangle
=\frac{1}{2}(\left|\beta_a\right\rangle\left|\beta_b\right\rangle+\left|-\beta_a\right\rangle\left|\beta_b\right\rangle+\left|\beta_a\right\rangle
\left|-\beta_b\right\rangle-\left|-\beta_a\right\rangle\left|-\beta_b\right\rangle),\nonumber \\
\end{eqnarray}
\end{small}
where $\beta_a=\alpha_a\exp{[ig^2\pi/(\chi\delta_a)]}$. Note that the entangled coherent states have many applications
in the field of quantum information. For instance, they can be used to construct
quantum gates \cite{s48} (using coherent states as the logical qubits \cite{s49}), implement quantum key
distribution \cite{s50}, build quantum repeaters \cite{s51}, test violation of Bell inequalities \cite{s52,s53},
and applicant in quantum metrology \cite{s54}.

\section{Possible experimental implementation}
\label{sec4}

When the inter-resonator crosstalk is taken
into account, the Hamiltonian~(2) becomes
$\widetilde{H}_{I}=H_{I}+\epsilon$, where $\epsilon$ describes the unwanted inter-resonator crosstalk,
given by $\epsilon=g_{ab}( e^{i\Delta_{ab} t}ab^{+}+h.c.)$, with the inter-resonator crosstalk coupling strength
$g_{ab}$ and the two-resonator frequency detuning $\Delta_{ab} =\omega_{a}-\omega_{b}$.

Including the dissipation and the dephasing, the dynamics of the lossy system is determined by the
following master equation
\begin{eqnarray}
\frac{d\rho }{dt} &=&-i[ \widetilde{H}_{I},\rho ] +\kappa _{a}
\mathcal{L}[a]+\kappa _{b}\mathcal{L}[b]\nonumber \\
&+&\gamma _{eg}\mathcal{L}[ \sigma
_{eg}^{-}] +\gamma _{fe}\mathcal{L}[ \sigma _{fe}^{-}]+\gamma _{fg}\mathcal{L}[ \sigma _{fg}^{-}]  \nonumber\\
&+&\sum_{j=e,f}\left\{ \gamma _{\varphi j}\left( \sigma _{jj}\rho
\sigma _{jj}-\sigma _{jj}\rho /2-\rho \sigma _{jj}/2\right) \right\},
\end{eqnarray}
where $\widetilde{H}_{I}$ is given above, $\sigma _{eg}^{-}=\left\vert g\right\rangle\left\langle
e\right\vert$, $\sigma _{fe}^{-}=\left\vert e\right\rangle \left\langle
f\right\vert$, $\sigma _{fg}^{-}=\left\vert g\right\rangle \left\langle
f\right\vert , \sigma _{jj}=\left\vert j\right\rangle\left\langle
j\right\vert (j=e,f);$ and $\mathcal{L}\left[ \Lambda \right] =\Lambda \rho \Lambda
^{+}-\Lambda ^{+}\Lambda \rho /2-\rho \Lambda ^{+}\Lambda /2,$ with $\Lambda=a,b,\sigma _{eg}^{-},\sigma _{fe}^{-},\sigma _{fg}^{-}.$ Here, $\kappa
_{a}$ ($\kappa_{b}$) is the photon decay rate of resonator $a$ ($b$). In addition, $
\gamma _{eg}$ is the energy relaxation rate of the level $\left\vert
e\right\rangle $ of qutrit, $\gamma _{fe}$ ($\gamma _{fg}$) is the
energy relaxation rate of the level $\left\vert f\right\rangle $ of qutrit for the decay path $\left\vert f\right\rangle \rightarrow \left\vert
e\right\rangle $ ($\left\vert g\right\rangle $), and $\gamma _{\varphi j}$ is the dephasing rate of the level $\left\vert
j\right\rangle $ of qutrit ($j=e,f$).

The fidelity of the operation is given by
\begin{eqnarray}
\mathcal{F}=\sqrt{\left\langle \psi _{\mathrm{id}}\right\vert \rho
\left\vert \psi _{\mathrm{id}}\right\rangle},
\end{eqnarray}
where $\left\vert \psi _{\mathrm{id}}\right\rangle $ is the output state of an ideal system (i.e., without dissipation, dephasing, and
crosstalk); while $\rho$ is the final density operator of the system when the operation is performed in
a realistic situation.

\subsection{Fidelity for the two-resonator qubits controlled-phase gate}
As an example, we will consider the case of $\alpha=\beta=\gamma=\delta=1/\sqrt{2}$. In this case, we have the qutrit-resonator-system initially
state $|\varphi\rangle_a$$|\varphi\rangle_b\otimes|g\rangle=(1/2)(|0\rangle_a+|1\rangle_a)(|0\rangle_b+|1\rangle_b)\otimes|g\rangle$
and the output state $\left\vert \psi _{\mathrm{id}}\right\rangle=|\phi _{\mathrm{id}}\rangle_{ab}\otimes|g\rangle$. Here, $|\phi _{\mathrm{id}}\rangle_{ab}$
is given by $|\phi _{\mathrm{id}}\rangle_{ab}
=(1/2)(|0\rangle_a|0\rangle_b+|0\rangle_a|1\rangle_b+|1\rangle_a|0\rangle_b
-|1\rangle_a|1\rangle_b)$, which is obtained based on Eq.~(8) for $|\theta| t=\pi$.
\begin{figure}[tbp]
\begin{center}
\includegraphics[bb=0 1 482 313, width=8.5 cm, clip]{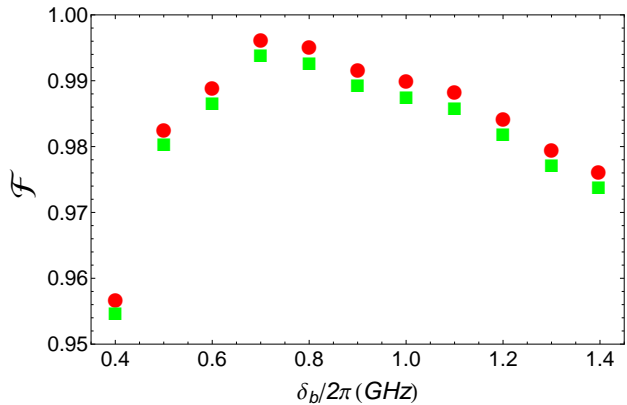}
\vspace*{-0.08in}
\end{center}
\caption{ Fidelity $\mathcal{F}$ versus detuning $\delta _{b}/2\pi$. Red dots correspond to the case without
considering decoherence of qutrit and dissipation of resonators; while green squares
correspond to the case that the systematic dissipation and dephasing are taken into account. The parameters used in the numerical simulation are referred to the text.}
\label{fig:}
\end{figure}

By solving the master equation (16), the fidelity of the gate operation can be calculated based on Eq~(17).
We now numerically calculate the fidelity for the operation.
The frequency of SC resonators typically is 1 to 10 GHz. Thus, we choose $\omega_{a}/2\pi=3.5$ GHz and $\omega _{b}/2\pi=6.5$ GHz such that $\Delta_{ab}=-3.0$ GHz. We then choose $\delta_a/2\pi=-0.3$ GHz and $g/2\pi=50$ MHz in our scheme. In the following,
we set the inter-resonator crosstalk strength $g_{ab}=0.1g$, which can be readily met in experiments \cite{12yang}.
Furthermore, we set $k$=1, $\gamma^{-1} _{j,\varphi e}=\gamma ^{-1}_{j,\varphi f}=\gamma$,
$\gamma^{-1} _{eg}=3\gamma,\gamma ^{-1}_{fe}=2\gamma$, $\gamma ^{-1}_{fg}=10\gamma$, and $\kappa_a^{-1}=\kappa_b^{-1}=\eta$.

Fig.~2 shows the fidelity versus $\delta _{b}/2\pi$. Green squares are plotted by choosing $\gamma=10~\mu s $ and $\eta=20~\mu s$, which
correspond to the case that the systematic dissipation and dephasing are taken into account. Here we consider a rather conservative case for the decoherence times of flux qutrit~\cite{Pop,s13}. It is noted that by designing the flux qutrit,
the $|g\rangle \leftrightarrow |f\rangle $ dipole matrix element can be made much smaller than that of the
$|g\rangle \leftrightarrow |e\rangle $ and $|e\rangle \leftrightarrow |f\rangle $ transitions. Thus,
$\gamma _{fg}^{-1}\gg \gamma _{eg}^{-1}, \gamma_{fe}^{-1}$. In addition, the leakage errors of a SC qutrit have been efficiently reduced in experiments \cite{Z. J. Chen}.

From Fig.~2, one can see that for $\delta _{b}/2\pi
=0.7$ GHz, a high fidelity $\sim$99.4\% is achievable for a two-resonator qubits controlled-phase gate.
For $\delta _{b}/2\pi=0.7$ GHz, one has $\mu/2\pi\sim$ 342 MHz.  The values of $g$ and $\mu$ are readily available in experiments \cite{10Niemczyk}.
In addition, one can obtain $\omega_{eg}/2\pi=3.2$ GHz and $\omega _{fe }/2\pi=7.2$ GHz. The transition frequency between two neighbor levels of a SC flux qutrit is typically in the range of 1-30 GHz. With the above given parameters, we obtain the large cross-Kerr interaction coefficient $\chi\sim 4.2$ MHz. Our numerical simulation indicates that the
high-fidelity implementation of a controlled-phase gate on two resonators is feasible
with current circuit QED technology. As in Fig. 2, the effect of the qutrit decoherence  and resonator decay on the fidelity is very small with the current parameter values.
To illustrate the effect of the dissipation and dephasing of the system,
we employ shorter qutrit decoherence and resonator-decay times in Fig.~3.

Fig.~3 displays the fidelity versus $\gamma$ and $\eta$, which is plotted by
choosing $\delta _{b}/2\pi=0.7$ GHz. From Fig.~4, one
can obtain $\{{\mathcal{F}},\gamma,\eta\}$: (i) $\{0.949, 0.1~\mu s, 1.0~\mu s\}$; (ii) $\{0.972, 0.3~\mu s, 2.0~\mu s\}$; (iii) $\{0.983, 0.5~\mu s, 3.0~\mu s\}$; (iv) $\{0.987, 0.7~\mu s, 4.0~\mu s\}$; and (v) $\{0.991, 1.0~\mu s, 5.0~\mu s\}$. Fig.~3 shows that for $\gamma \in[0.1\mu s,1\mu s ]$ and $\eta \in[1\mu s ,5\mu s ]$, the fidelity can be greater than $94\%$. This is because the qutrit is unexcited and resonator photons are virtually excited during the entire process, qutrit decoherence and resonator decay can be efficiently suppressed.

\begin{figure}[tbp]
\begin{center}
\includegraphics[bb=13 397 522 759, width=8.5 cm, clip]{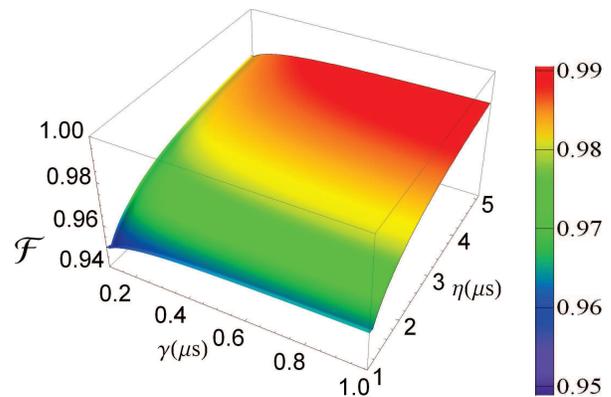}
\vspace*{-0.08in}
\end{center}
\caption{ Fidelity versus $\gamma$ and $\eta$, which is plotted by
choosing $\delta _{b}/2\pi=0.7$ GHz.}
\label{fig:3}
\end{figure}

For the resonator frequencies and the resonator-decay times used in the Fig.~2, the required quality factors of the two resonators are $Q_a\sim4.4\times10^{5}$ and $Q_b\sim8.2\times10^{5}$, which are attainable in experiments because a
quality factor $Q\sim10^{6}$ for SC coplanar waveguide resonators have been experimentally demonstrated \cite{s40,s41}.
For Fig.~3, the required quality factors for the resonators are $Q_a\sim[2.2\times10^{4},1.1\times10^{5}]$ and $Q_b\sim[4.1\times10^{4},2.0\times10^{5}]$.
Fig.~3 shows that the phase-gate operation also can be high-fidelity performed assisted by the low-$Q$ resonators.

\subsection{Fidelity for generation of a two-resonator entangled coherent state}

The fidelity for the operation is calculated based on Eq~(17), where the
ideal output state is $\left\vert \psi _{\mathrm{id}}\right\rangle=|\varphi\rangle\otimes|g\rangle$
and $\rho$ is obtained by numerically
solving the master equation~(16) for an initial input state $|\varphi\rangle_a|\varphi\rangle_b\otimes|g\rangle=$ $|\alpha_a\rangle|\beta_b\rangle\otimes|g\rangle$. Here, state $|\varphi\rangle$ is given by  Eq. (15).

We choose $\omega_{a}/2\pi=3.5$ GHz, $\omega _{b}/2\pi=6.5$ GHz, and $\delta_a/2\pi=-1.0$ GHz.
We set $g/2\pi=150$ MHz and $\mu/2\pi=200$ MHz because the coupling strengths of
values $g$ and $\mu$ are readily achievable in experiments \cite{10Niemczyk}.
In addition, we set $\gamma= 0.1~\mu s$ and  $\eta=5~\mu s$. The qutrit decoherence and the resonator-decay times
used here are referred to the Fig.~3.

\begin{figure}[tbp]
\begin{center}
\includegraphics[bb=-5 3 809 518, width=9.0 cm, clip]{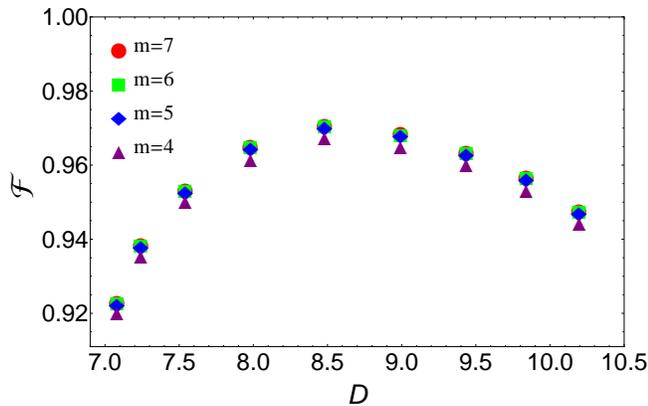}
\vspace*{-0.08in}
\end{center}
\caption{Fidelity versus the normalized detuning $D=\delta_b/\mu$, which is plotted by $m=4,5,6,7$.}
\label{fig:4}
\end{figure}

We now numerically calculate the fidelity for creation of the two-resonator entangled coherent state.
In our numerical calculation, we consider the first $m$ terms in the
expansions of coherent states $|\alpha_a\rangle$ ($|\beta_b\rangle$) and $|-\alpha_a\rangle$ ($|-\beta_b\rangle$). Figure~4 shows the fidelity versus the normalized detuning $D=\delta_b/\mu$ with $|\alpha_a|=0.5$ and $|\beta_b|=1.0$, which is plotted for $m=4,5,6,7$.
For $D=8.48$, a high fidelity 96.75 \%, 97.03 \%, 97.08 \%, 97.09 \% can be obtained for $m=$ 4, 5, 6, and 7, respectively.
For $D=8.48$, we have $\delta_b\sim~1.7$ GHz and  $\chi\sim 0.83$ MHz.

For resonators $a$ and $b$ with frequencies and dissipation times used
in the Fig.~4, the quality factors of the two resonators are $Q_a\sim 1.1\times10^{5}$ and $Q_b\sim 2.0\times10^{5}$.
The numerical simulation indicates that the
high-fidelity generation of a two-resonator entangled coherent state is feasible
with current circuit QED technology.

\section{Conclusion}
\label{sec5}

We have proposed a method for realizing the cross-Kerr nonlinear interaction
between two microwave resonators induced by a superconducting flux qutrit.
Our present proposal differs from the previous protocols~\cite{Y. Hu,H. Zhang}.
First, in our proposal only one qutrit is needed, thus the circuit complexity is much reduced.
Second, there is only need one operation step and unnecessary to employ classical pulse, so the
operation procedure is greatly simplified.
Finally, due to the resonator photons are virtually excited and the coupler is unexcited for
the entire process, the effect of resonator decay, the unwanted inter-resonator crosstalk, and the
coupler decoherence are greatly minimized.

Although we assume that the cross-Kerr nonlinearity effect is performed
between two SC coplanar waveguide resonators, using the three-level
flux qutrit, our proposal can in principle also be applied
to other solid devices, for example, the schemes based on other kinds of SC qutrits
(e.g., SC charge qutrits, transmon qutrits, Xmon
qutrits, phase qutrits) coupled to
two 1D resonators or two 3D SC cavities, or based on the two nitrogen-vacancy center ensembles (behaves as two bosonic modes) coupled to a flux qutrit.

Based on our cross-Kerr interaction Hamiltonian, we implement a two-resonator qubits controlled-phase gate and generate a two-resonator entangled coherent state. Numerical simulation shows that
the high-fidelity implementation of the
phase gate and creation of the entangled coherent state are feasible with state-of-the-art circuit
QED technology.
Our finding provides a new way for realizing the cross-Kerr nonlinearity interaction between two microwave resonators, and such cross-Kerr effect may find applications in quantum information processing.

\section*{ACKNOWLEDGEMENTS}
This work was supported by the Ministry of Science and
Technology of China under Grant No. 2016YFA0301802,
the National Natural Science
Foundation of China under Grants No. 11375036 and No.
11175033, the Xinghai Scholar Cultivation Plan, and the
Fundamental Research Funds for the Central Universities
under Grants No. DUT15LK35 and No. DUT15TD47.

\end{document}